\documentclass[aps,prd,preprint,groupedaddress,showpacs]{revtex4-1}

\usepackage{amsmath,amssymb,graphicx}
\usepackage{tikz}
\usetikzlibrary{decorations.pathmorphing,snakes}

\begin{document}


\title{Critical Behavior in a Massless Scalar Field Collapse \\ with Self-interaction Potential}


\author{Xuefeng Zhang}
\email[]{zhxf@bnu.edu.cn}
\affiliation{Department of Physics, Department of Astronomy, Beijing Normal University, Beijing 100875, China}

\author{H. L\"u}
\email[]{mrhonglu@gmail.com}
\affiliation{Department of Physics, Beijing Normal University, Beijing 100875, China}


\date{\today}

\begin{abstract}
  We examine a one-parameter family of analytical solutions representing spherically symmetric collapse of a nonlinear massless scalar field with self-interaction in an asymptotically flat spacetime. The time evolution exhibits a type of critical behavior. Depending on the scalar charge parameter $q$ as compared to a critical value $q^*$, the incoming scalar wave collapses either to a globally naked central singularity if $q<q^*$ (weak field) or to a scalar-hairy black hole if $q>q^*$ (strong field), both having finite asymptotic masses. Near the critical evolution, the black hole mass follows a product-logarithmic scaling law: $-M^2\ln M \sim q-q^*$ with $0<M\ll 1$ and $q>q^*$. The solution admits no self-similarity and satisfies the null and the strong energy conditions.
\end{abstract}


\maketitle

\section{Introduction}

Sufficient mass accumulation in a confined space forms a black hole. Due to the stability of the Minkowski spacetime, sufficiently small initial data will remain small in time evolution. Both scenarios were explored analytically by Christodoulou \cite{Christo}, for which he considered minimally coupled massless scalar fields in spherical symmetry, and established various fundamental theorems. Using the same model, Choptuik investigated the threshold of black hole formation by numerical means \cite{Choptuik93}, and demonstrated that the scalar wave either collapses to a black hole or disperses to infinity, with a critical solution separating them. Universality and self-similarity of the critical solution and power scaling laws of the black hole masses were identified, which have now fallen under the name ``critical phenomena'' (see, e.g., review articles \cite{Wang01, Gundlach07}). Since Choptuik's breakthrough, critical phenomena have been extended to many other matter fields. Particularly for minimally coupled scalar fields with potentials, they have been observed in quadratic potentials $m^2\phi^2$ (massive scalar) \cite{Brady97} and symmetrical double-well potentials \cite{Honda02}.

Besides the numerical approach, attempts have been made to explore the possibility of constructing exact solutions that exhibits critical behaviors in the Einstein-scalar theory. Unfortunately, such solutions seem quite rare and difficult to find. Perhaps the most well-known example is Roberts' one-parameter self-similar solution \cite{Roberts89}, which has been analysed for different parameter regimes \cite{Brady94} and used as a toy model for critical collapse. The Wyman solution has also been argued to exhibit a type of critical behavior \cite{Oliveira05}, even though it is static and has no significant relevance to gravitational collapse.

Very recently, a one-parameter family of exact time-dependent spherically symmetric solutions were discovered in four-dimensional Einstein gravity minimally coupled to a dilaton scalar field with a self-interaction potential \cite{Zhang14}. For a certain range of the parameter, the solution represents gravitational collapse to a static scalar-hairy black hole in an (A)dS or Minkowski background. The solution also reduces to the Roberts solution in an appropriate limit, and therefore a connection to critical collapse was suspected. In this paper, we will show that this suspicion is justified.

The paper is organized as follows. In the next section, we will first present the time-dependent solution along with its scalar invariants, the energy-momentum tensor and masses. The energy conditions and the relation to the Roberts solution are also discussed. In Section \ref{sec_crit}, we move on to details of critical behaviors in three distinct parameter regimes: the subcritical, critical and supercritical. We use the Misner-Sharp mass to identify the apparent horizon, the causal nature of the central singularity and the black hole mass. Additionally, we analyze outgoing radial null geodesics and draw the Penrose diagrams for each regime. Then the paper is concluded in Section \ref{sec_conc}.

\section{The time-dependent solution} \label{sec_soln}

In this section, we consider the minimally coupled Einstein-scalar field with a scalar potential. The Lagrangian density takes the form
\begin{eqnarray}
 \mathcal{L} &=& \sqrt{-g} \left[R - \frac{1}{2}\left(\nabla_\alpha \phi \nabla^\alpha \phi\right) - V(\phi)\right], \label{L} \\
 V(\phi) &=& 4\lambda \left[3\sinh\phi - \phi(\cosh\phi + 2)\right], \label{V}
\end{eqnarray}
where $\phi$ is a real-valued scalar field and $V$ its potential with a coupling constant $\lambda$. This particular potential function was due to Zloshchastiev \cite{Zlosh05} (see also \cite{Gonzalez13}), for which we have removed terms that contributes to a nonzero cosmological constant such that $V(0)=0$. The potential has an odd parity and a Taylor expansion $V(\phi)=-\frac{\lambda}{15}\phi^5-\frac{\lambda}{315}\phi^7+\cdots$ at the origin which accounts for a quintic interaction in the leading order term. The mass of the scalar field is defined as $m_\phi^2=V''|_{\phi=0}=0$. Hence we are dealing with a self-coupled nonlinear massless scalar field. Moreover, the potential is strictly monotonic and unbounded from below (asymptotically, $V(\phi)\sim -2\lambda\phi\exp(\pm\phi)$ for $\phi\rightarrow \pm\infty$). Therefore various established no-hair theorems for minimally coupled scalar fields can be bypassed (see, e.g., \cite{Bekenstein72}).
\begin{figure}[htbp]
 \includegraphics[width=0.5\textwidth]{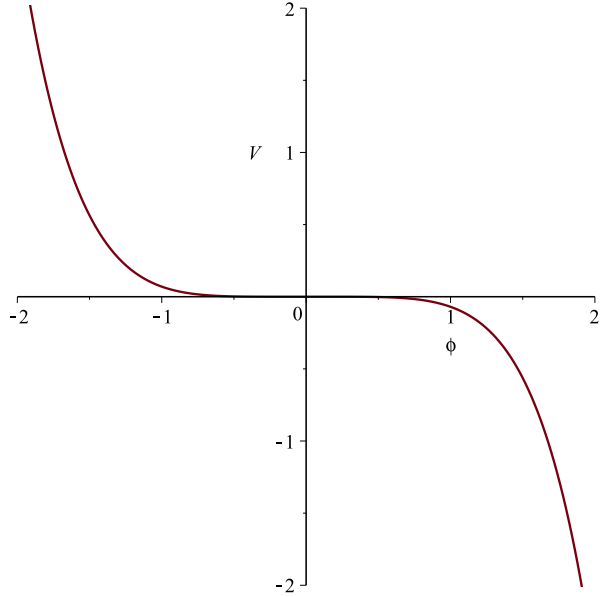}
 \caption{The scalar potential $V(\phi)$ given by (\ref{V}) with $\lambda=1$. \label{Vphi}}
\end{figure}

From the Lagrangian density (\ref{L}), the energy-momentum tensor for the scalar field is
\begin{equation} \label{T}
 2T_{\mu\nu} = \nabla_\mu \phi \nabla_\nu \phi - \left[\frac{1}{2}\left(\nabla_\alpha \phi \nabla^\alpha \phi\right) + V(\phi)\right] g_{\mu\nu}.
\end{equation}
The Einstein equations ($G_{\mu\nu}=T_{\mu\nu}$) and the Klein-Gordon equation can be written as
\begin{eqnarray}
 R_{\mu\nu} &=& \frac{1}{2}\big(\nabla_\mu \phi \nabla_\nu \phi + V(\phi)g_{\mu\nu}\big), \\
 \nabla_\alpha \nabla^\alpha \phi &=& V'(\phi).
\end{eqnarray}
To construct time-dependent solutions, we have followed the Newman-Penrose formalism and further extended a generalized form of the Robinson-Trautman solution \cite{Guven96, Stephani03} for nonlinear scalar fields. In Eddington-Finkelstein-like coordinates, the resulting metric \cite{Zhang14} is given by
\begin{eqnarray}
 \mathrm{d}s^2 = 2 \mathrm{d}v\mathrm{d}r - H(r,v)\mathrm{d}v^2 + R^2(r,v)\mathrm{d}\Omega^2, \label{metricr}\\
 \phi(r,v) = \ln\!\left(1 + \frac{q}{r} \tanh(\lambda qv)\right),
 \qquad R^2 = r^2\left(1 + \frac{q}{r}\tanh(\lambda qv)\right), \\
 H = 1 - \lambda q^2 - 2\lambda q r \tanh(\lambda qv)
 + 2\lambda r^2 \left(1 + \frac{q}{r}\tanh(\lambda qv)\right)
 \ln\!\left(1 + \frac{q}{r}\tanh(\lambda qv)\right), \label{H}
\end{eqnarray}
where $v$ is an advanced time and the area radius $R$ is not to be confused with the Ricci scalar $R^\alpha_{\phantom{\alpha}\alpha}$. The solution contains one free real parameter $q$, commonly known as the scalar charge. For the Newman-Penrose formalism, one can pick the following the null tetrad:
\begin{eqnarray}
 \mathrm{d}s^2=2\omega^1\omega^2-2\omega^3\omega^4,
 \qquad \mathrm{d}\Omega^2=\frac{2\mathrm{d}\zeta\mathrm{d}\bar\zeta}{(1+\frac{1}{2}\zeta\bar\zeta)^2}
 = \mathrm{d}\theta^2+\sin^2\theta\mathrm{d}\varphi^2, \\
 \omega^1 = \frac{R(r,v)}{1+\frac{1}{2}\zeta\bar\zeta} \mathrm{d}\zeta = \bar\omega^2,
 \qquad \omega^3 = -\mathrm{d}v, \qquad \omega^4 = \mathrm{d}r - \tfrac{1}{2}H(r,v)\mathrm{d}v.
\end{eqnarray}
with the complex coordinate $\zeta = \sqrt{2}\tan(\theta/2)\exp(\mathrm{i}\varphi)$. Accordingly, the only non-vanishing Weyl scalar is
\begin{equation}
 \Psi_2 = -\frac{\lambda q^3\tanh(\lambda qv)}{6r^2\left(r+q\tanh(\lambda qv)\right)^2}
 \left(r-\frac{1-\lambda q^2}{2\lambda q}\tanh(\lambda qv)\right).
\end{equation}
Thereby the spacetime is of the Petrov type D, and it possesses two congruences of null geodesics along the double principal null directions $\partial_r$ and $2\partial_v+H\partial_r$, both being shearfree and non-twisting. The Ricci scalar is
\begin{eqnarray}
 R^\alpha_{\phantom{\alpha}\alpha} &=&
 \left(\partial_u\phi+\tfrac{1}{2}H\partial_r\phi\right)\partial_r\phi + 2V(\phi) \nonumber \\
 &=& -\frac{3\lambda\left(8r^2+8qr\tanh(\lambda qv)+q^2\tanh^2(\lambda qv)\right)}{r(r+q\tanh(\lambda qv))}
 \ln\!\left(1+\frac{q}{r}\tanh(\lambda qv)\right) \nonumber \\
 & & + \frac{q\tanh(\lambda qv)}{2r^2(r+q\tanh(\lambda qv))^2}
 \Big[48\lambda r^3+72\lambda qr^2\tanh(\lambda qv) \nonumber \\
 & & -2\lambda q^2 r\left(1-12\tanh^2(\lambda qv)\right)
 +(1-\lambda q^2)q\tanh(\lambda qv)\Big]. \label{Ricci}
\end{eqnarray}
Both quantities indicate a curvature singularity at the center $r=0$ and that the metric is asymptotically flat as $r\rightarrow +\infty$.

To study the global structure, it is preferable to use the area radius $R$ for coordinates (though we will still perform certain calculations with the $(r,v)$-coordinates for simplicity). Thus with
\begin{equation}
 r = \sqrt{R^2+\tfrac{1}{4}q^2\tanh^2(\lambda qv)}-\tfrac{1}{2}q\tanh(\lambda qv), \label{rR}
\end{equation}
we can transform the metric into a standard form for spherically symmetric spacetimes:
\begin{eqnarray}
 \mathrm{d}s^2 = 2\mathrm{e}^{\beta(R,v)} \mathrm{d}v\mathrm{d}R -
 \mathrm{e}^{2\beta(R,v)}\left(1-\frac{2M(R,v)}{R}\right)\mathrm{d}v^2 + R^2\mathrm{d}\Omega^2,
 \label{metricR} \\
 \phi = \ln\!\left(\frac{\sqrt{R^2+\frac{1}{4}q^2\tanh^2(\lambda qv)}+\frac{1}{2}q\tanh(\lambda qv)}
 {\sqrt{R^2+\frac{1}{4}q^2\tanh^2(\lambda qv)}-\frac{1}{2}q\tanh(\lambda qv)}\right),
 \mathrm{e}^\beta = \frac{R}{\sqrt{R^2+\frac{1}{4}q^2\tanh^2(\lambda qv)}}, \\
 \mathrm{e}^{2\beta}\left(1-\frac{2M}{R}\right) =
 1-\frac{2\lambda q(R^2+\frac{1}{4}q^2)\tanh(\lambda qv)}{\sqrt{R^2+\frac{1}{4}q^2\tanh^2(\lambda qv)}} \nonumber \\
 +2\lambda R^2\ln\!\left(\frac{\sqrt{R^2+\frac{1}{4}q^2\tanh^2(\lambda qv)}+\frac{1}{2}q\tanh(\lambda qv)}
 {\sqrt{R^2+\frac{1}{4}q^2\tanh^2(\lambda qv)}-\frac{1}{2}q\tanh(\lambda qv)}\right), \label{HR}
\end{eqnarray}
where the function $M(R,v)=R(1 - g^{\mu\nu} \partial_\mu R \partial_\nu R)/2$ is the Misner-Sharp quasi-local mass which one may interpret as the gravitational mass inside a sphere of the area radius $R$ at an advanced time $v$. This mass function reduces to the Bondi-Sachs (BS) and ADM masses in appropriate limits \cite{Hayward96}. Specifically, one has
\begin{equation} \label{MBS}
 M_{BS}(v) = \lim_{R\rightarrow +\infty} M(R,v)
 = \tfrac{1}{12}\lambda q^3 \tanh(\lambda qv) \left(3-\tanh^2(\lambda qv) \right),
\end{equation}
which measures the mass on null hypersurfaces of constant $v$. Similarly, the ADM mass is
\begin{equation} \label{MADM}
 M_{ADM} = \lim_{v\rightarrow +\infty} M_{BS}(v) = \tfrac{1}{6} \lambda q^3,
\end{equation}
which represents the conserved total mass of the scalar field.

To ensure the positivity of the asymptotic masses, one should at least require $\lambda q\geq 0$. Without loss of generality, we only consider $\lambda>0$ and $q>0$ from now on. The case for $\lambda<0$ and $q<0$ can be equally obtained by mapping $\lambda\rightarrow -\lambda$, $q\rightarrow -q$ and $\phi\rightarrow -\phi$ which leaves $V(\phi)$, the equations of motion and the metric (\ref{metricR}) invariant.

In terms of the coordinate $R$, the Ricci scalar, as a curvature invariant, can be calculated by directly plugging (\ref{rR}) into (\ref{Ricci}). Particularly near $v=0$, we have
\begin{equation}
 R^{\alpha}_{\phantom\alpha\alpha} = -\frac{\lambda^2 q^4}{R^3}v+\mathcal{O}(v^2).
\end{equation}
Expressions for the energy-momentum tensor are, unfortunately, too cumbersome to be fully exhibited here:
\begin{eqnarray}
 T_{RR} = \frac{q^2\tanh^2(\lambda qv)}{2R^2\left(R^2+\frac{1}{4}q^2\tanh^2(\lambda qv)\right)}
 = \frac{q^4\lambda^2}{2R^4}v^2+\mathcal{O}(v^3), \\
 T_{Rv} = -\frac{\lambda^2 q^4}{4R^4}v^2+\mathcal{O}(v^3), \qquad
 T_{vv} = \frac{\lambda^2 q^4}{2R^2}-\frac{\lambda^2 q^4}{2R^3}v+\mathcal{O}(v^2), \\
 T_{\theta\theta} = \frac{\lambda^2 q^4}{2R}v+\mathcal{O}(v^2), \qquad
 T_{\varphi\varphi} = \frac{\lambda^2 q^4\sin^2\theta}{2R}v+\mathcal{O}(v^2).
\end{eqnarray}
We make a note that all listed components above become singular at $R=0$ (analogous to $T_{\mu\nu}$ of the Vaidya solution \cite{Vaidya51}) and that the only non-vanishing component at $v=0$ is $T_{vv}$.

The energy-momentum tensor (\ref{T}) for minimally coupled scalar fields with potentials always satisfies the null energy condition. Additionally, since $V(\phi)\leq 0$, it satisfies the strong energy condition as well \cite{Barcelo00}. However, the weak energy conditions can be violated locally in the spacetime, depending on the parameters $\lambda$ and $q$. But we can show that it is fulfilled for large $r$ in the static limit (see Appendix \ref{app_A}).

Furthermore, the solution we have described so far is closely related to the Roberts solution \cite{Roberts89} for linear scalar fields without self-interaction potentials. To recover the latter, one only needs to reparameterize $q=\sqrt{p/\lambda}$ and take $\lambda\rightarrow 0$ such that $V(\phi)$ vanishes and the metric (\ref{metricr}) becomes
\begin{equation}
 \mathrm{d}s^2 = 2\mathrm{d}v\mathrm{d}r - (1-p)\mathrm{d}v^2 + r(r + pv)\mathrm{d}\Omega^2,
 \qquad \phi=\ln\!\left(1+\frac{pv}{r}\right),\qquad p>0.
\end{equation}
Thereby we may follow Roberts' example \cite{Brady94} in discussing various global aspects of our solution. Comparisons between the two may also benefit our discussion. In this regard, we recall that the black hole mass of the Roberts solution grows to infinity as $v\rightarrow +\infty$ (to see this, perform the aforementioned limit in (\ref{MADM})). This undesirable property is clearly avoided in our solution on account of the extra $V(\phi)$.

\section{Critical behavior} \label{sec_crit}

Given that the quantities $\Psi_2$, $R^\alpha_{\phantom\alpha\alpha}$ and $M(R,v)$, as well as the scalar field $\phi$, all vanish at $v=0$, we can match the spacetime continuously across the null hypersurface $v=0$ with a Minkowski spacetime for $v\leq 0$. Thus we consider incoming flux of the scalar wave to be turned on at the advanced time $v=0$. By doing so, we also exclude the region where the Bondi-Sachs mass becomes negative (cf. (\ref{MBS}) with $v<0$).

As $v\rightarrow +\infty$, we have $\tanh(\lambda qv)\rightarrow 1$ and the solution (\ref{metricr}) (also (\ref{metricR})) reaches a static limit. With the time-dependence fading away, we can rewrite the resulting metric in a Schwarzschild-like form \cite{Gonzalez13}:
\begin{eqnarray}
 \mathrm{d}s^2 = -f(r)\mathrm{d}t^2+\frac{\mathrm{d}r^2}{f(r)}+R^2(r)\mathrm{d}\Omega^2,
 \qquad \phi(r) = 1+\frac{q}{r}, \label{static} \\
 R^2=r(r+q), \qquad
 f = \lim_{v\rightarrow +\infty}\!\!H = 1 - \lambda q^2 - 2\lambda qr + 2\lambda r^2\left(1+\frac{q}{r}\right)\ln\!\left(1+\frac{q}{r}\right),
\end{eqnarray}
where we have adopted $\mathrm{d}t=\mathrm{d}v+\mathrm{d}r/f$. To reinstate the area radius $R$, one can resort to
\begin{equation}
 r = \sqrt{R^2+\tfrac{1}{4}q^2} - \tfrac{1}{2}q,
 \qquad \mathrm{d}r = \frac{R}{\sqrt{R^2+\frac{1}{4}q^2}}\mathrm{d}R.
\end{equation}
Analogous to the Wyman solution \cite{Oliveira05}, the static solution itself also demonstrates a ``phase transition'' at the special value $q=1/\sqrt{\lambda}$. A first clue for this is to look at the outgoing null geodesics emanated from the center:
\begin{equation}
 2\frac{\mathrm{d}r}{\mathrm{d}v}\,\bigg|_{r\rightarrow 0+}
 \!\!= f(r)\big|_{r\rightarrow 0+} \!\!= 1-\lambda q^2,
\end{equation}
which indicates that the singularity is completely censored (no light emitted) if $1-\lambda q^2<0$, but (at least) locally naked if $1-\lambda q^2>0$. This relates to the fact that the central singularity is covered by an event horizon only when $1-\lambda q^2<0$ \cite{Gonzalez13}.

\begin{figure}[htbp]
  \includegraphics[width=0.6\textwidth]{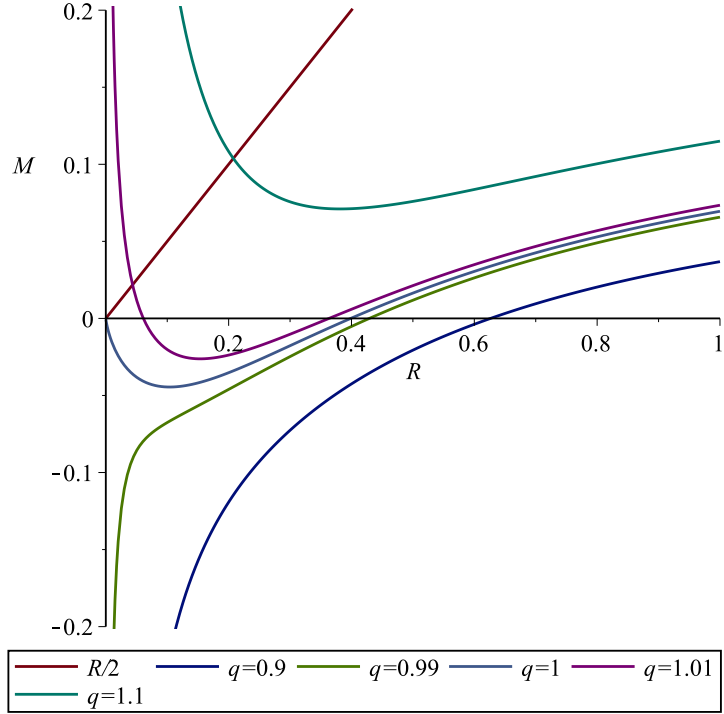}
  \caption{The Misner-Sharp mass function $M(R,v)$ in the static limit $v\rightarrow +\infty$ and with $\lambda=1$. The critical case corresponds to $q=1$. Event horizons are located at intersections with $R/2$.}
  \label{fig_MS1}
\end{figure}

Similarly for the dynamical solution, the existence of an apparent horizon also relies on the sign of $1-\lambda q^2$. The apparent horizon is determined by the root of $g^{\mu\nu}\partial_\mu R\partial_\nu R = 1-2M(R,v)/R=0$. For a given $v>0$ (also recall $\lambda>0$, $q>0$), the metric function $\mathrm{e}^{2\beta}(1-2M/R)$ is strictly increasing from $1-\lambda q^2$ to $1$ as $R$ goes from $0$ to $+\infty$. Hence a unique apparent horizon exists iff $1-\lambda q^2<0$. Furthermore, the criticality of $q=1/\sqrt{\lambda}$ also presents itself in the Misner-Sharp mass function:
\begin{eqnarray}
 M(R,v) = -\frac{1-\lambda q^2}{8R}\,q^2\tanh^2(\lambda qv) + \mathcal{O}(R\ln R), \\
 \Longrightarrow \ \ \ \
 \lim_{R\rightarrow 0+} M(R, v>0) =
 \begin{cases}
 -\infty, & q<1/\sqrt{\lambda}, \\
 0, & q=1/\sqrt{\lambda}, \\
 +\infty, & q>1/\sqrt{\lambda}. \\
 \end{cases} \label{MR0}
\end{eqnarray}
The sign of $M$ above implies that the central singularity is timelike and untrapped for $q<1/\sqrt{\lambda}$, and spacelike and trapped for $q>1/\sqrt{\lambda}$ \cite{Hayward96}. Typical graphs of the mass function in the static limit are given in FIG. \ref{fig_MS1}. From the plot, one can see that the mass function is typically increasing for large $R$. But for $q>1/\sqrt{\lambda}$, it can be decreasing immediately outside the event horizon, which indicates a local breakdown of the dominant energy condition \cite{Hayward96}.

Depending on the value of the parameter $q$, we can now classify the time-dependent solution into three different types, i.e., the subcritical ($0<q<1/\sqrt{\lambda}$), critical ($q=1/\sqrt{\lambda}$) and supercritical ($q>1/\sqrt{\lambda}$).

\subsection{Subcritical $\big(0<q<1/\sqrt{\lambda}\big)$}

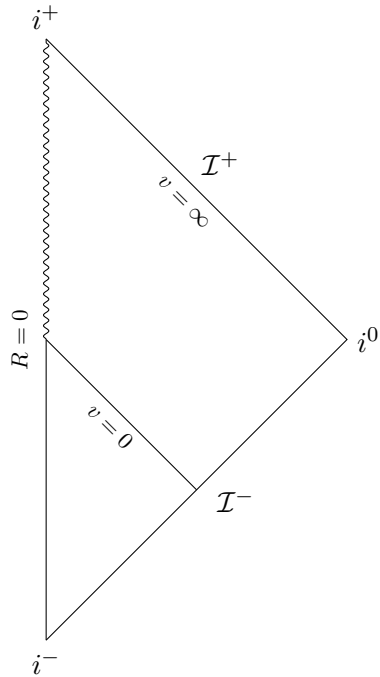
\begin{figure}[htbp]
  \centering
  \begin{tikzpicture}[scale=4]
    \draw (0,0) node[below]{$i^-$} -- node[midway, below=4pt, right=4pt]{$\mathcal{I}^-$}
       (1,1) node[right]{$i^0$} -- node[pos=0.55, above=4pt, right=4pt]{$\mathcal{I}^+$}
                                   node[midway, below=2pt, sloped, scale=0.8]{$v=\infty$}
       (0,2) node[above]{$i^+$};
    \draw (0,0) -- (0,1) node[at end, above=4pt, sloped, scale=0.8]{$R=0$};
    \draw (0.5,0.5) -- (0,1) node[midway, below, sloped, scale=0.8]{$v=0$};
    \draw[snake=snake, segment amplitude=1pt, segment length=4pt] (0,1) -- (0,2);
  \end{tikzpicture}
  \caption{The Penrose diagram for subcritical evolutions with $0<q<1/\sqrt{\lambda}$. The incoming scalar wave from the past null infinity $\mathcal{I}^-$ is turned on at $v=0$ before which the spacetime is flat and empty. A timelike naked singularity at $R=0$ forms and persists as the scalar wave collides at the center. }
  \label{fig_pen1}
\end{figure}

The Penrose diagram is given in FIG. \ref{fig_pen1}. The solution represents a scalar field collapsing from the past null infinity towards the center at $R=0$. The gravitational interaction is not strong enough to create a black hole. However, the wave packet is not dispersing away to infinity after colliding at the center. Instead, it forms a globally naked timelike singularity which has negative mass and becomes static as the time progresses. This collapse outcome is a major deviation from the subcritical case of the Roberts solution and Choptuik's numerical examples \cite{Choptuik93}. Naked central singularities have also been shown to exist in self-similar collapse of scalar fields \cite{Brady95}.

\subsection{Critical $(q=1/\sqrt{\lambda})$}

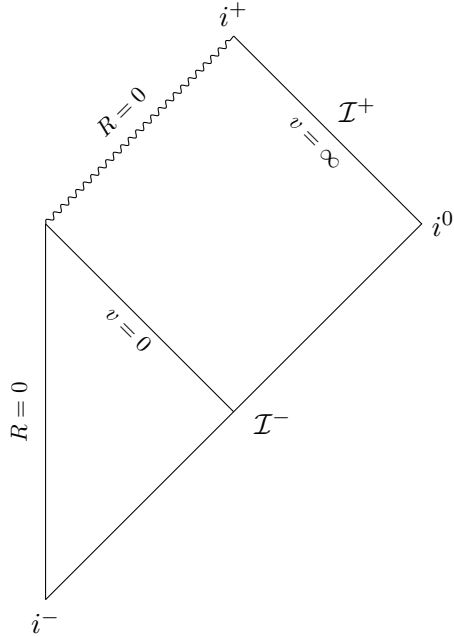
\begin{figure}[htbp]
  \begin{tikzpicture}[scale=5]
   \draw (0,0) node[below]{$i^-$} -- node[midway, below=4pt, right=4pt]{$\mathcal{I}^-$}
         (1,1) node[right]{$i^0$} -- node[pos=0.55, above=4pt, right=4pt]{$\mathcal{I}^+$}
                                   node[midway, below=2pt, sloped, scale=0.8]{$v=\infty$}
         (0.5,1.5) node[above]{$i^+$};
   \draw (0,0) -- (0,1) node[midway, above=4pt, sloped, scale=0.8]{$R=0$};
   \draw (0.5,0.5) -- (0,1) node[midway, below, sloped, scale=0.8]{$v=0$};
   \draw[snake=snake, segment amplitude=1pt, segment length=4pt]
       (0,1) -- (0.5,1.5) node[midway, above=4pt, sloped, scale=0.8]{$R=0$};
  \end{tikzpicture}
  \caption{The Penrose diagram for the critical evolution with $q=1/\sqrt{\lambda}$. The initial configuration is similar to the subcritical case. The collapse now proceeds to a null massless singularity at $R=0$. }
  \label{fig_pen2}
\end{figure}

The Penrose diagram is shown in FIG. \ref{fig_pen2}, which is qualitatively no different from the critical case of the Roberts solution \cite{Brady94}. The central singularity is also null due to
\begin{equation}
 \lim_{R\rightarrow 0+} \mathrm{e}^{2\beta}\left(1-\frac{2M}{R}\right) = 1-\lambda q^2=0.
\end{equation}
The mass function vanishes at the singularity (cf. (\ref{MR0})), while the curvature diverges there. In the $(r,v)$-coordinates, the outgoing radial null geodesics obey
\begin{equation} \label{geo}
 2\,\frac{\mathrm{d}r}{\mathrm{d}v} = H(r,v), \qquad v\geq 0.
\end{equation}
At the center, we have
\begin{equation}
 \lim_{r\rightarrow 0+} H(r,v\geq 0) = 1-\lambda q^2 = 0.
\end{equation}
Hence the above geodesic equation has a constant solution $r(v\geq 0)=0$ (i.e., $R=0$). In Appendix \ref{app_B}, we prove the local uniqueness of this constant solution, which means that no null geodesics initiated at $r(v_0)=0$ with $v_0\geq 0$ can ever leave $r=0$, let alone reaching the null infinity. For outgoing null rays with $r(v_0)>0$, they will always reach the null infinity since the function $H(r>0,v>0)$ is positive and strictly increasing in both $r$ and $v$. In summary, there is no apparent horizon in the spacetime; observers can detect signals from an arbitrarily small vicinity (with arbitrarily large curvature) of the central singularity, but no direct signals from it until they actually run into the singularity. Hence one may think of this critical spacetime as an intermediate state between a naked singularity and a black hole.

It is normally expected that critical solutions in scalar field collapse may possess self-similarity, e.g., the Roberts solution. However, in Appendix \ref{app_C}, we show that self-similarity (homothety) is in fact missing from our solution, not only for the critical case, but for all $q>0$.

\subsection{Supercritical $(q>1/\sqrt{\lambda})$}

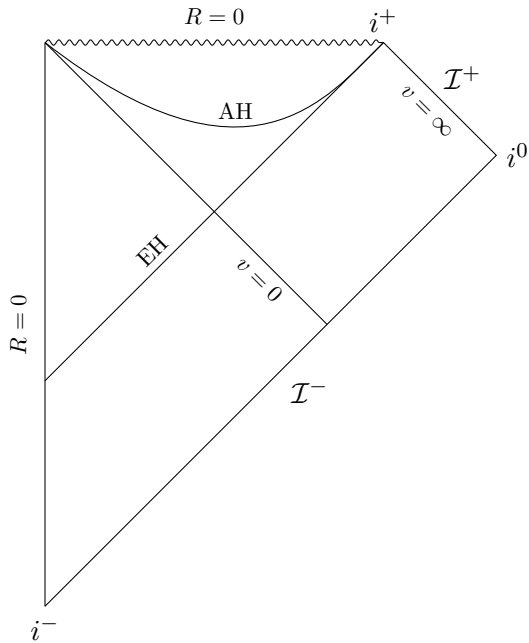
\begin{figure}[htbp]
  \begin{tikzpicture}[scale=6]
   \draw (0,0) node[below]{$i^-$} -- node[midway, below=4pt, right=4pt]{$\mathcal{I}^-$}
         (1,1) node[right]{$i^0$} -- node[pos=0.55, above=6pt, right]{$\mathcal{I}^+$}
                                   node[midway, below=2pt, sloped, scale=0.8]{$v=\infty$}
         (0.75,1.25) node[above]{$i^+$};
   \draw (0,0) -- (0,1.25) node[midway, above=4pt, sloped, scale=0.8]{$R=0$};
   \draw (0.625,0.625) -- (0,1.25) node[pos=0.2, below, sloped, scale=0.8]{$v=0$};
   \draw[snake=snake, segment amplitude=1pt, segment length=4pt]
         (0,1.25) -- (0.75,1.25) node[midway, above=4pt, scale=0.8]{$R=0$};
   \draw (0,0.5) -- (0.75,1.25) node[pos=0.35, above, sloped, scale=0.8]{EH};
   \draw (0,1.25) .. controls (0.45,0.9) and (0.625,1.125) .. (0.75,1.25) node[pos=0.4, above, scale=0.8]{AH};
  \end{tikzpicture}
  \caption{The Penrose diagram for supercritical evolutions with $q>1/\sqrt{\lambda}$. The influx of the scalar field $(v\geq 0)$ generates a black hole which contains a spacelike singularity at the center. The apparent horizon (AH) is spacelike and approaches the event horizon (EH) from the inside as $v\rightarrow +\infty$.}
  \label{fig_pen3}
\end{figure}

\begin{figure}[htbp]
  \includegraphics[width=0.5\textwidth]{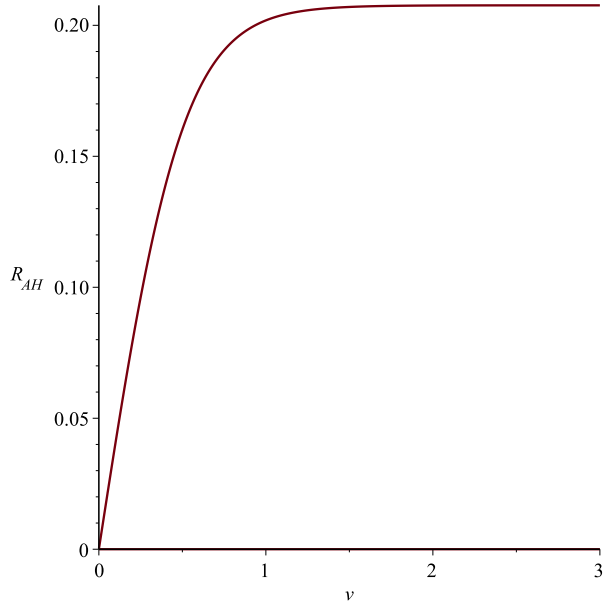}
  \caption{Numerical determination of the apparent horizon $R_{AH}(v)$ with $\lambda=1$ and $q=1.1$. As $v$ grows, the curve reaches a plateau that corresponds to the event horizon. }
  \label{fig_AH}
\end{figure}

This case captures strong-field collapse of the scalar field. FIG. \ref{fig_pen3} shows the global structure of the spacetime. The central singularity at $R=0, v\geq 0$ is now spacelike, completely censored and surrounded by an apparent horizon. As mentioned before, one needs to solve $M(R,v)=R/2$ for $R=R_{AH}(v)$ to determine the location of the apparent horizon as a function of $v$. This can be done numerically, and in FIG. \ref{fig_AH}, we give an example of $R_{AH}(v)$ plotted with Maple 16 (\verb"implicitplot"). As a consequence of the null energy condition, the spacelike apparent horizon $R_{AH}$ should be monotonically increasing in $v$ \cite{Hayward96} and tend to a finite value as it approaches the event horizon from the inside. This is indeed confirmed by the numerical plot, as well as all other samples tested by us with various inputs of $q$.

By definition, the event horizon is always null. To locate the event horizon, one needs to look for outgoing radial null geodesics that are not able to reach infinity by propagating them from the center of the spacetime. Preferably, this can be done through numerical integration, since analytical solutions are very unlikely to obtain. As one may expect, the event horizon extends beyond the null hypersurface $v=0$ into the initial flat region, which means that it comes to exist before the central singularity starts to form. This is a teleological feature also seen in the Vaidya spacetime. Moreover, in the limit $v\rightarrow +\infty$, the event horizon of the dynamical black hole will match the static one (cf. (\ref{static})).

In dynamical spacetimes, a black hole can be locally defined by its apparent horizon. Thereby one can evaluate the Misner-Sharp mass on the apparent horizon as a measure of the mass of a dynamical black hole, i.e., $M_{AH}(v) = R_{AH}(v)/2$. Following this definition, in order to obtain the final mass of the black hole, one needs to take the limit $v\rightarrow +\infty$:
\begin{equation}
 M_F = R_F/2, \qquad R_F = \lim_{v\rightarrow +\infty} R_{AH}(v).
\end{equation}
Here the final area radius of the black hole, denoted by $R_F$, is determined by
\begin{equation}
 0 = 1-\lambda q \sqrt{4R_F^2+q^2} + 2\lambda R_F^2\ln\!\left(\frac{\sqrt{4R_F^2+q^2}+q} {\sqrt{4R_F^2+q^2}-q}\right).
\end{equation}
Now consider the near-critical regime with $q\simeq 1/\sqrt{\lambda}$. For a small $R_F$, the right-hand side of the above equation has the following generalized series expansion:
\begin{equation}
 0 = 1-\lambda q^2-4\lambda R_F^2\ln R_F + \mathcal{O}(R_F^2).
\end{equation}
Ignoring higher order terms, we obtain an approximate scaling law for the black hole mass:
\begin{equation}
 -M_F^2\ln M_F \sim q - q^*, \qquad q^*=1/\sqrt{\lambda}.
\end{equation}
This product-logarithmic relation (Lambert $W$-function) is significantly different from the power law of Choptuik. In addition, by a similar argument for small $v>0$, we have also determined that the initial growth of the black hole mass follows
\begin{equation}
 M_{AH} \sim (q - q^*)^{1/2} v,
\end{equation}
which resembles the power law of the Roberts solution for having the same exponent $1/2$.

\section{Conclusion} \label{sec_conc}

The basic physical picture we have drawn so far is the following: an imploding scalar wave packet, with the spherical wave front (located at the advanced time $v=0$) travelling inward at the speed of light, collides at the center of the 3-dimensional physical space. Depending on the strength of the wave packet, it either forms a globally naked singularity in weak-field collapse, or a hairy black hole in strong-field collapse. In case of the latter, the formation of black holes always starts with zero mass. In both cases, the spacetime is becoming static as the collapse reaches its ending. Meanwhile, the total mass stays finite. There also exists a critical case interpolating between these two dominant outcomes, of which the spacetime is sitting on the verge of containing a naked singularity or a black hole, but just avoiding both.

The phenomenology of our model has unusual features, but it still fits into the basic picture of critical collapse. Major deviations from previously known examples of critical phenomena are at least threefold: 1) the presence of naked singularities as the endstate of weak-field collapse instead of dispersal to infinity, 2) the absence of self-similarity in the critical case, 3) the product-logarithmic scaling law of the black hole mass in contrast to the usual power law. All these features may be attributed to the unbounded negative scalar potential. With the black hole mass being kept finite, our solution can also be considered as a ``regularized'' generalization of the Roberts solution which has been thought unfavorable for not having an event horizon.

For future work, we point out that more is yet to be understood regarding the legitimacy and universality of the ``critical solution'' with $q=1/\sqrt{\lambda}$. The stability of the solution also raises important questions. Furthermore, our earlier paper \cite{Zhang14} also includes dS and AdS generalizations of the solution (\ref{metricr}). It is expected that the local picture of the naked singularity/black hole formation may not be affected by the introduction of an effective cosmological constant, at least when the constant is small. Nonetheless, the global structures will be quite different. Perhaps more interestingly, the singularity/black hole ``phase transition'' in the AdS background may be further examined in the context of the AdS/CFT correspondence.

\appendix
\section{The weak energy condition} \label{app_A}

Without loss of generality, we consider a normalized timelike geodesic vector field $v^\mu=\mathrm{d}x^\mu/\mathrm{d}\tau$ such that $v^\alpha v_\alpha=-1$ and $v^\mu\nabla_\mu v^\nu=0$. The weak energy condition can be expressed as
\begin{equation}
 0 \leq 2T_{\mu\nu} v^\mu v^\nu
 = (\phi')^2 + \tfrac{1}{2}\left(\nabla_\alpha \phi \nabla^\alpha \phi\right) + V(\phi),
\end{equation}
with $\phi'=\mathrm{d}\phi/\mathrm{d}\tau=v^\mu\nabla_\mu \phi$. The inequality holds true if the sum of the last two terms above is non-negative. Given the solution (\ref{metricr}), we can estimate them as follows:
\begin{eqnarray}
 \tfrac{1}{2}\left(\nabla_\alpha \phi \nabla^\alpha \phi\right) + V(\phi)
 &=& \left(\partial_v\phi+\tfrac{1}{2}H\partial_r\phi\right)\partial_r\phi + V(\phi) \\
 &=& -\tanh(\lambda qv)(1-\tanh^2(\lambda qv)) \frac{\lambda q^3}{r^3} \nonumber \\
 & & + \tanh^2(\lambda qv)\left[1+3\lambda q^2\big(1-\tanh^2(\lambda qv)\big)\right]\frac{q^2}{2r^4}
 + \mathcal{O}(r^{-5}),
\end{eqnarray}
where the second equality stands for a series expansion at $r=+\infty$. Now consider the static limit $v\rightarrow +\infty$ in which case the (negative) leading term vanishes. The sum is thus dominated by $q^2/2r^4>0$ for large $r$. So the weak energy condition at least holds in this situation.

Another noteworthy situation is to consider a small neighborhood of the center $r=0$, where we have
\begin{eqnarray}
 & &\tfrac{1}{2}\left(\nabla_\alpha \phi \nabla^\alpha \phi\right) + V(\phi) \nonumber \\
 &=& \frac{1-\lambda q^2}{2r^2} + \frac{1-6\lambda q^2\tanh^2(\lambda qv)+\lambda q^2\tanh^2(\lambda qv)\ln(q\tanh(\lambda qv) r^{-1})}{q\tanh(\lambda qv)r} + \mathcal{O}(1).
\end{eqnarray}
The leading term indicates that for $v>0$ and $0<q<1/\sqrt{\lambda}$, the weak energy condition is satisfied near the timelike central singularity in subcritical evolutions.

\section{Outgoing null geodesics $(q=1/\sqrt{\lambda})$} \label{app_B}

In this section, we prove the (local) uniqueness of the constant solution $r(v\geq 0)=0$ for the null geodesic equation (\ref{geo}). Because $\partial H/\partial r$ blows up at $r=0$, the Picard-Lindel\"{o}f theorem fails to apply. Instead we will use the comparison theorem (generalized Gronwall's inequality) for nonlinear first-order ODEs (\cite{Walter98}, see \S 9). Now consider a ``backward'' initial value problem (IVP):
\begin{equation} \label{IVP}
 \frac{\mathrm{d}r}{\mathrm{d}v} = \frac{H(r,v)}{2}, \qquad r(v_0)=r_0>0, \qquad 0\leq v \leq v_0.
\end{equation}
where $H$, given by (\ref{H}), is non-negative. All we need to show is that the solution to this IVP can never reach $r=0$ with $0\leq v \leq v_0$, i.e., $r(v)>0$.

For the critical case $\lambda=1/q^2$, the right-hand side of the equation is bounded above by
\begin{eqnarray}
 \frac{q^2}{2} H(r,v)
 &=& r\left(r+q\tanh(q^{-1}v)\right)\ln\!\left(1+\frac{q\tanh(q^{-1}v)}{r}\right) - rq\tanh(q^{-1}v)
 \nonumber \\
 &\leq& r\left(r+A\right)\ln\!\left(1+\frac{A}{r}\right) - Ar
 < r\left(r+A\right)\ln\!\left(1+\frac{A}{r}\right),
\end{eqnarray}
with $A=q\tanh(q^{-1}v_0)>0$. Here we have used the fact that the function $H(r,v)$ is monotonically increasing in $v$. The bounding differential equation
\begin{equation}
 q^2 \frac{\mathrm{d}\tilde{r}}{\mathrm{d}v} = \tilde{r}\left(\tilde{r}+A\right)\ln\!\left(1+\frac{A}{\tilde{r}}\right)
\end{equation}
can be solved analytically and the general solution is
\begin{equation}
 \tilde{r}(v) = \frac{A}{\mathrm{e}^{C\mathrm{e}^{-Av/q^2}}-1}.
\end{equation}
with an integration constant $C$. Particularly for the initial condition $\tilde{r}(v_0)=r_0$, the constant $C$ can be determined as
\begin{equation}
 C = \mathrm{e}^{Av_0/q^2}\ln\!\left(1+\frac{A}{r_0}\right)>0,
\end{equation}
so that we have $\tilde{r}(v)>0$. Then by the virtue of the comparison theorem, the solution of (\ref{IVP}) is bounded below by
\begin{equation}
 r(v) \geq \tilde{r}(v) > 0, \qquad 0\leq v \leq v_0.
\end{equation}
This completes the proof.

\section{Absence of self-similarity} \label{app_C}

In this section, we show that the dynamical solution (\ref{metricr}) is not self-similar, i.e., that it does not admit a homothetic Killing vector $\xi=\xi^r(r,v)\partial_r + \xi^v(r,v)\partial_v$ satisfying
\begin{equation}
 \xi_{\mu;\nu} + \xi_{\mu;\nu} = A g_{\mu\nu}
\end{equation}
with $A$ a non-zero real constant. To see this, we first write down explicitly components of the above equation:
\begin{eqnarray}
 \partial_r \xi^v &=& 0, \label{KE1} \\
 \tfrac{1}{2}\left(\partial_v\xi^v + \partial_r\xi^r\right) &=& A, \label{KE2} \\
 -H\partial_v\xi^v
 -\tfrac{1}{2}(\partial_v H)\xi^v + \partial_v\xi^r - \tfrac{1}{2}(\partial_r H)\xi^r &=& -AH, \label{KE3} \\
 \tfrac{1}{2}\lambda q^2 r\left(1-\tanh^2(\lambda qv)\right)\xi^v
 + \left(r+\tfrac{1}{2}q\tanh(\lambda qv)\right)\xi^r &=& Ar\big(r+q\tanh(\lambda qv)\big). \label{KE4}
\end{eqnarray}
Note that the first two equations imply that $\partial_r^2 \xi^r=0$, which, when combined with (\ref{KE4}), yields
\begin{equation}
 \xi^v = \frac{A\tanh(\lambda qv)}{\lambda q(1-\tanh^2(\lambda qv))}.
\end{equation}
Then again from (\ref{KE4}), one obtains $\xi^r=Ar$. Plugging them back in (\ref{KE2}) gives one
\begin{equation}
 A \cosh^2(\lambda qv) = A,
\end{equation}
which fails to hold if $A\neq 0$. Therefore we conclude that a homothetic vector does not exist for our solution.

\begin{acknowledgments}
 We are grateful to Xinliang An, Daniel Finley and Tong-Jie Zhang for useful discussions and kind assistance. We thank for the questions and comments from anonymous referees of our paper \cite{Zhang14}, which this follow-up paper is intended to address. The work is funded by China Postdoctoral Science Foundation and in part by the NSFC grants 11175269 and 11235003.
\end{acknowledgments}


\end{document}